\documentclass[aps,prb,twocolumn,floatfix,floats,showpacs,superscriptaddress,raggedbottom]{revtex4}
\usepackage{graphicx,latexsym}
\usepackage{dcolumn}
\usepackage{amsmath}

\def\fig#1{Fig.~\ref{#1}}

\begin{document}

\title{Coherent magnetotransport spectroscopy in an edge-blocked double quantum
wire\\ with window and resonator coupling}

\author{Chi-Shung Tang}
\affiliation{Physics Division, National Center for Theoretical
        Sciences, P.O.\ Box 2-131, Hsinchu 30013, Taiwan}

\author{Vidar Gudmundsson}
\affiliation{Science Institute, University of Iceland,
        Dunhaga 3, IS-107 Reykjavik, Iceland}

%

\begin{abstract}
We propose an electronic double quantum wire system that contains a
pair of edge blocking potential and a coupling element in the middle
barrier between two ballistic quantum wires.  A window and a
resonator coupling control between the parallel wires are discussed
and compared for the enhancement of the interwire transfer processes
in an appropriate magnetic field.  We illustrate the results of the
analysis by performing computational simulations on the conductance
and probability density of electron waves in the window and
resonator coupled double wire system.
\end{abstract}



\maketitle

%
%

\section{Introduction}

Coherent transport spectroscopy allows us to explore localized
resonance processes when states interact through a coupling element.
Earlier experimental considerations include the tunneling transfer
of states between electron waveguides through a thin tunneling
barrier,\cite{Eugster91} and the window coupling between diffusive
wires.\cite{Hirayama92}  Later on, theoretical studies were
performed on noninteger conductance steps in a gapped double
waveguide,\cite{Xu95} and tunneling conductance between two coupled
waveguides.\cite{Governale00} Very recently, Rashba spin-orbit
effect in parallel quantum wires has also been
studied.\cite{Guzenko06}  Of particular interest are the dynamics of
the transfer processes for single-energy electron spectroscopy in
coupled quantum states with either tunneling\cite{Tang05} or window
coupling.\cite{Gudmundsson06} However, the optimal transfer
conditions in the coupled electronic systems have not been
investigated.

In the presence of magnetic field, the energy spectra have been
studied pointing out the complex structure of the evanescent states
of the system in a homogeneous double wire,\cite{Barbosa97}  in a
disordered double wire with boundary
roughness,\cite{Korepov02,Arapan03} and in spatially coincident
coupled electron waveguides.\cite{Fischer05}  Recently,
magnetotransport in a parallel double wire coupled through a
potential barrier was studied by Shi and Gu.\cite{Shi97} They have
shown that the stepwise conductance increasing and decreasing
features can be changed by the applied magnetic field and the height
of the potential barrier.

\begin{figure}[htbq]
      \includegraphics[width=0.45\textwidth,angle=0]{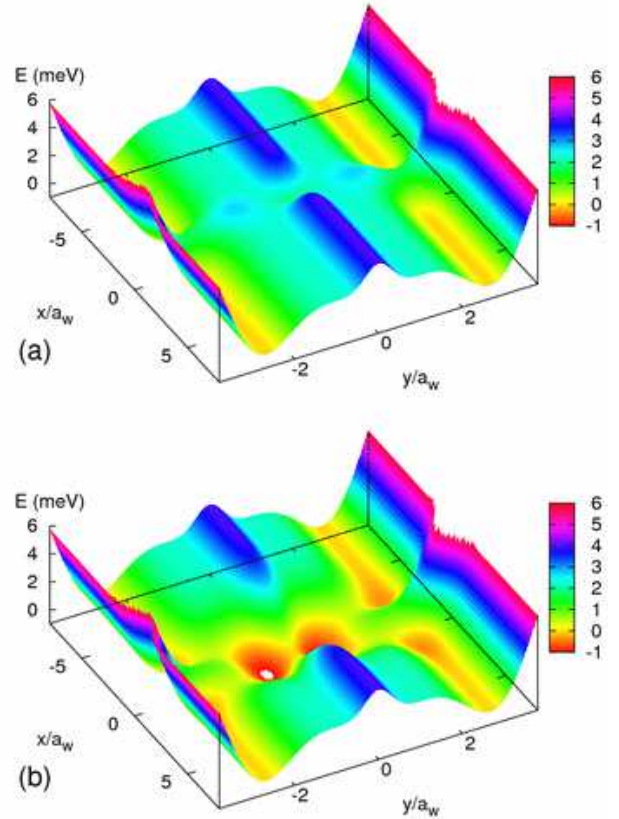}
      \caption{(Color online) Schematic illustration of edge-blocked
      double quantum wire with (a) window coupling and (b) resonator coupling
      between the wires.  The color scale on the right shows the potential height in meV.
      The effective confinement length $a_w = 33.7$ nm in zero magnetic field.}
      \label{System}
\end{figure}

In this paper, we investigate how a resonator in the coupling window
and a uniform perpendicular magnetic field affect the electronic
transport characteristics in a ballistic two-terminal double wire
system. In order to enhance the interwire transfer coupling, a pair
of edge-blocking potentials are also included. The potential
landscapes of the controlled coupled systems are depicted in
\fig{System}.

We shall demonstrate coherent magnetotransport properties in an edge
blocked double wire system using a rigorous Lippmann-Schwinger
formalism in a momentum-coordinate space,\cite{Gurvitz95} and
transforming the two-dimensional magnetotransport equation into a
set of coupled one-dimensional integral equations for the
$T$-matrix.\cite{Gudmundsson05}  The characteristics in conductance
and electron probability density of the system show the dynamics of
forward and backward interwire transfer by properly adjusting the
strength of the magnetic field, the window size, and the resonator.

%
\section{Edge-Blocked Double Wire}
The system under investigation is composed of a laterally parallel
double quantum wire with transverse confining potential
\begin{equation}
      V_{\rm conf}(y) = \frac{1}{2}m^*\Omega_0^2 y^2 +
      V_{\delta}(y) + V_{00}
\label{V_conf}
\end{equation}
with a symmetric deviation
\begin{align}
      V_{\delta}(y)=&- V_{d_1} \exp{\left[-\beta_1 (y-y_0)^2\right]}\nonumber\\
                    &+ V_{d_0} \exp{(-\beta_0  y^2        )}\nonumber\\
                    &- V_{d_1} \exp{\left[-\beta_1(y+y_0)^2\right]}
\label{V_delta}
\end{align}
for the sake of forming a parallel double wire from the parabolic
confinement, in which $V_{00}$ denotes a global shift of potential
to avoid negative energy in the wire. The typical parameters for the
confining potentials have the values: $\hbar\Omega_0 = 1.0$ meV,
$V_{00}=2.0$ meV, $V_{d_0} = 2.0$ meV, $V_{d_1} = 6.0$ meV, $\beta_0
= 4.0\times 10^{-3}$ nm$^{-2}$, $\beta_1 = 7.0\times 10^{-4}$
nm$^{-2}$, and $y_0 = 100$ nm.

The scattering potential
\begin{equation}
      V_{\rm sc}(x,y) =  V_{\rm block}(x,y) + V_{\rm coup}(x,y)
\label{Vsc}
\end{equation}
contains two terms, namely an edge blocking potential
\begin{equation}
      V_{\rm block}(x,y) = V_1\exp{(-\gamma_x x^2)}\left[ 1- \exp{(-\gamma_y
      y^2)}\right] ,
\label{Vblock}
\end{equation}
and a coupling element
\begin{equation}
      V_{\rm coup}(x,y) = V_0\exp{(-\alpha x^2-\beta y^2)}.
\label{Vcouple}
\end{equation}
These scattering potentials could be implemented in an experimental
system by means of depositing Schottky front gates.\cite{Sasaki06}
Based on the scattering potentials presented above we discuss two
different coupling elements: (i) the edge blocking with simple
window coupling if $V_0 = -V_{d_0}$, and (ii) the edge blocking with
resonator coupling if $V_0 = -2 V_{d_0}$. First, with the choice
$V_0 = -V_{d_0}$, $\alpha =\alpha_{\rm W}= 2.0\times 10^{-4}$
nm$^{-2}$, and $\beta = \beta_0$, a simple window coupling between
the parallel wires can be made with the edge blocking strength $V_1
= 6$ meV. Further, with the choice $V_0 = -2 V_{d_0}$, $\alpha
=\alpha_{\rm R} =0.2\alpha_{\rm W}$, and $\beta = 0.05\beta_0$, a
window resonator coupling between the parallel wires can be made
with the increased edge blocking strength $V_1 = 8$ meV. For both
cases, the other parameters for the edge blocking potential are:
$\gamma_x = \beta_1$ and $\gamma_y = \gamma_x / 15.9$.

Using a momentum-coordinate representation,\cite{Gurvitz95} the
lateral component Hamiltonian for the conduction electrons in a
double wire system can be written in the form
\begin{eqnarray}
H(q,y) &=& \left[\frac{p_y^2}{2m^*} +
\frac{1}{2}m^*\Omega_w^2(y-y_q)^2 \right.\nonumber \\
&& + \left. \frac{\hbar^2q^2}{2m^*} \left( 1-
\frac{\omega_c^2}{\Omega_w^2} \right) + V_{\delta}(y) \right],
\end{eqnarray}
where $y_q = qa_w^2 \omega_c/\Omega_w$ is the effective parabolic
center shift.  The effective magnetic length
$a_w=\hbar/(m^*\Omega_w)$ with $\Omega_w^2=\Omega_0^2+\omega_c^2$
and $\omega_c=eB/(m^*c)$ being the two-dimensional cyclotron
frequency.  The wave functions in the considered double wire system
away from the scattering region can be generally written in the
expansion form
\begin{equation}
      \Psi_E(q,y) = \sum_n \varphi_n(q)\Phi_n(q,y)
\end{equation}
containing the eigenfunctions $\Phi_n(q,y)$ of the double wire
confinement Eq.\ (\ref{V_conf}), which can be expanded in terms of
the eigenfunctions for the parabolic confinement\cite{Gurvitz95}
\begin{equation}
      \Phi_n(q,y) = \sum_n c_{nm}(q)\phi_m(q,y),
      \label{Phinqy}
\end{equation}
where $\phi_m(q,y)$ is an eigenfunction for the parabolic wire with
finite magnetic field. The coefficients $c_{nm}(q)$ are independent
of the energy $E$ of the incident electron waves, and can be
obtained by diagonalizing separately the deviated Hamiltonian in
each $q$-subspace.  Then we reduce the Lippmann-Schwinger equation
into a set of coupled one-dimensional integral equations for the
$T$-matrix.\cite{Gudmundsson05}

The matrix elements of the scattering potential are of the integral
form
\begin{eqnarray}
      V_{nn'}(q,p) &=& \int dy \Phi_n^*(q,y) V_{\rm sc}(q-p,y) \Phi_{n'}(p,y) \nonumber \\
      &=& \sum_{ls}c_{nl}^*(q) c_{n's}(p) V_{ls}(q,p),
\end{eqnarray}
where $V_{ls}(q,p)$ is given by
\begin{equation}
      V_{ls}(q,p) = \int dy \phi_l^*(q,y)V(q-p,y)\phi_s(p,y).
\end{equation}
In the asymptotic regions of the double wire, the propagating
electrons can be described by
\begin{equation}
\Psi_E(q,y) = 2\pi \delta\left[q-k_n(E)\right]\Phi_n(q,y).
\end{equation}
The corresponding energy subbands $E_n(q)$ for the incoming
propagating states are represented by
\begin{equation}
      E_n(q) = E_n^0(q)  + \frac{(qa_w)^2}{2}
      \frac{(\hbar\Omega_0)^2}{\hbar\Omega_w},
\label{E_nq}
\end{equation}
where $E_n^0(q) =  E_n^0 + \epsilon (n,q)$ contains contributions
from the parabolic confinement $E_n^0=\hbar\Omega_w(n+1/2)$ and the
correction $\epsilon (n,q)$ due to the deviation potential
$V_{\delta}(y)$.   It should be noted that the deviation energy
$\epsilon (n,q)$ makes the subbands generally not equidistant in
energy.

To achieve numerical accuracy, we notice that the evanescent modes
are in general not orthogonal and centered around $y = 0$, whereas
the propagating modes shift along the $y$-direction and their number
of nodes correlates to the subband index $n$. This fact leads us to
expand the evanescent modes in terms of the {\it unshifted}
eigenfunctions, but for the case of finite magnetic field, namely
the complete basis
\begin{equation}
      \phi_n^0(y) = \frac{\exp{\left(-\frac{y^2}{2a_w}\right)}}{\sqrt{2^n\:\sqrt{\pi}
      \: n!\: a_w}}H_n\left(\frac{y}{a_w}\right) .
\label{phi_0}
\end{equation}
Using this basis and keeping the real part of the energy spectrum
for the evanescent modes, we find that the evanescent-mode energy
spectrum is consistent with the results of nonparabolic confinement
by Barbosa {\it et al.},\cite{Barbosa97} and Korepov {\it et
al}.\cite{Korepov02}  On the other hand, we would like to mention
that the character of the evanescent modes require a larger basis
than the expansion for propagating modes to obtain sufficient
numerical accuracy.

\begin{figure}[htbq]
      \includegraphics[width=0.45\textwidth,angle=0]{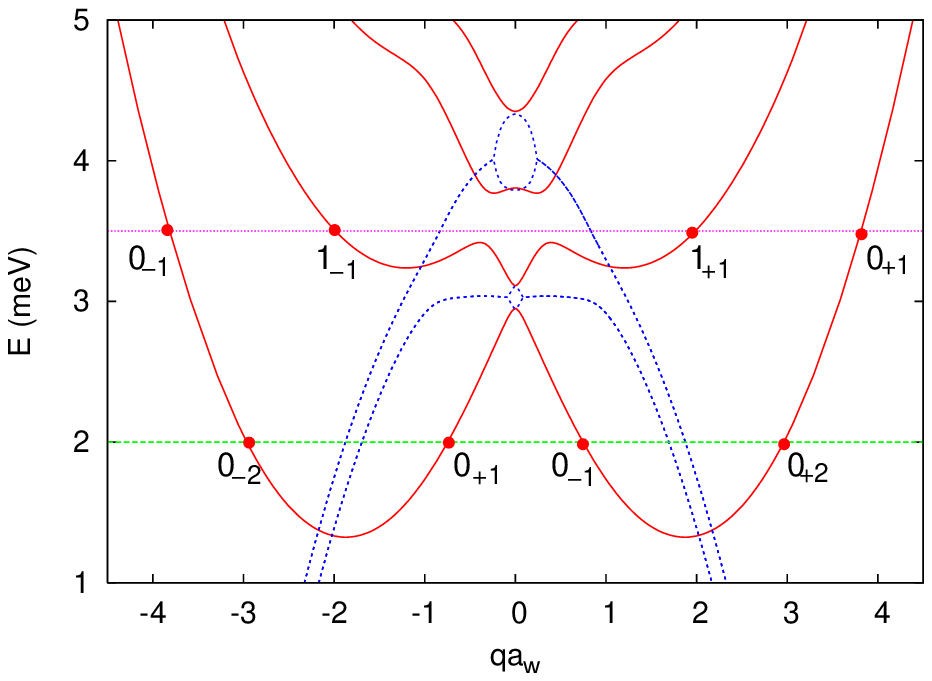}
      \caption{(Color online) The energy spectrum of the propagating
               electronic states (solid red) vs.\ the Fourier
               parameter $q$, and the energy spectrum of the evanescent
               states (dotted blue) vs.\ $iq$,
               in the double wire away from the coupling element with magnetic
               field $B=0.5$ T.
               The active transport modes are labeled for two energies
               with the notation $n_{\pm i}$, where the $+i$ and $-i$ indicate, respectively,
               the number of the right- and left-going active modes in the subband $n$.}
      \label{Eq}
\end{figure}

Due to the nonparabolic double wire confinement, each electron
subband provides more than one propagating mode corresponding to the
poles of the retarded scattering Green function
\begin{equation}
      G_E^n(q) = \left[(k_n(E)a_w)^2-(q a_w)^2 +
      i0^+\right]^{-1}.
\label{G_nE}
\end{equation}
Here $k_n(E)$ is independent of $q$, which is the Fourier variable
with no connection to $E$, and the index $i$ labels the modes in
subband $n$. The subband momentum $\hbar k_n(E)$ can be determined
by
\begin{equation}
      \left[k_n(E)a_w\right]^2 = 2\left[E-E_n^0(q)\right]
      \frac{\hbar\Omega_w}{(\hbar\Omega_0 )^2}.
\label{k_n}
\end{equation}
In contrast to the parabolic confinement measuring the kinetic
energy from the subband bottom at zero $q$, the electron kinetic
energy for a double wire system is measured from $E_n^0(q)$ with
finite $q$ values.  For an electron incident in mode $m_j$ with
momentum $\hbar k_{m_j}$, the transmission amplitude in mode $n_i$
with momentum $\hbar k_{n_i}$ is given by\cite{Gudmundsson05}
\begin{align}
      t_{{n_i}{m_j}}(E) = \delta_{{n_i}{m_j}} -&{\ }
      \frac{i\sqrt{(k_{m_j}/k_{n_i})}}{2(k_{m_j}a_w)}
      \left(\frac{\hbar\Omega_0}{\hbar\Omega_w} \right)^2\nonumber\\
      &\times {\tilde T}_{{n_i}{m_j}}(k_{n_i},k_{m_j}),
\label{t(E)}
\end{align}
where the positive subindexes $j$ and $i$ count, respectively, the
incident and forward scattering modes for a given subband index $m$
and $n$, as is illustrated in \fig{Eq}. Two different values of the
incoming energy are labeled for the active transport modes. The
conductance, according to the framework of Landauer-B{\"u}ttiker
formalism,\cite{Landauer57,Buttiker86} can be written as
\begin{equation}
      G(E) = G_0{\rm Tr}[ {\bf t}^{\dagger}(E){\bf t}(E)] ,
\label{G}
\end{equation}
where $G_0 = 2e^2/h$ and  ${\bf t}$ is evaluated at the Fermi
energy.
%

\section{Numerical Results and discussion}
To investigate the magnetotransport properties of the edge-blocked
double wire system with either window or resonator coupling, we
select a typical magnetic field strength $B=0.5$ T such that the
effective confinement length $a_w=29.3$ nm and the effective
parabolic subband separation $\hbar\Omega_w = 1.32$ meV.  We then
correlate the conductance for a certain incoming energy, and seek
information from the electron probability density of the various
modes active at that incoming energy. All the calculations presented
below are carried out under the assumption that the electrons have
an effective mass of $0.067m_e$, which is appropriate to the
AlGaAs/GaAs interface. Numerical accuracy is assured by comparing
the data obtained from a larger basis set.

In the absence of magnetic field, the deviation potential causes a
near degeneration between the lowest two subbands, and hence the
first plateau of the quantized conductance would be $2G_0$. However,
in the presence of magnetic field, the Lorentz force may push the
electrons away from the center of the system in ratio to their
longitudinal momenta and then destroy the parity of the electron
waves.  The lowest two subbands are no longer degenerate but form an
energy gap $\Delta_{01}=0.16$ meV between the $n=0$ subband top and
the $n=1$ subband bottom at $q=0$, shown in \fig{Eq}, reducing the
conductance to $G_0$.  For a homogeneous spatially separated double
wire without coupling element, the conductance plateaus are not
monotonically increased with increasing incident electron energy due
to the rich subband structure of the system.

\subsection{Window coupling}

\begin{figure}[bthq]
      \includegraphics[width=0.45\textwidth,angle=0]{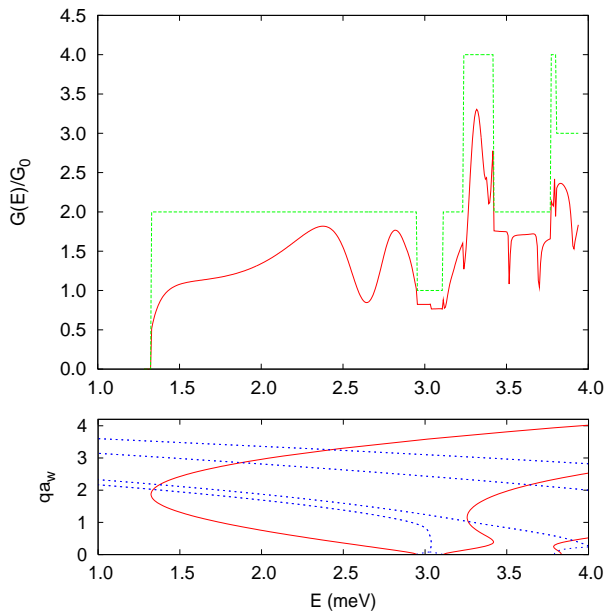}
      \caption{(Color online) Upper subfigure: Conductance of an
      edge-blocked double quantum wire without
               coupling element (dashed green) and with a simple window coupling between
               the wires (solid red).
               Lower subfigure: The energy spectrum of the propagating
               electronic states (solid red) vs.\ the Fourier
               parameter $q$, and the energy spectrum of the evanescent
               states (dotted blue) vs.\ $iq$,
               in the asymptotic regions of the system.
               $B=0.5$ T, $\alpha = 2.0\times 10^{-4}$ nm$^{-2}$, and $V_0 = -2.0$ meV.}
      \label{GE_W}
\end{figure}

To study the window coupled transport behavior of the edge-blocked
double wire, we select $\alpha_{\rm W}= 2.0\times 10^{-4}$ nm$^{-2}$
corresponding to the effective window length $L_{\rm W}=70$ nm, $V_0
= -V_{d_0}$, $\beta = \beta_0$, and the edge blocking strength $V_1
= 6$ meV.  The numerical results of the conductance spectroscopy and
its corresponding energy spectrum are depicted in \fig{GE_W}.

Without a coupling element, electrons with energy greater than the
pinch off energy $E=1.33$ meV, the double conductance step $G=2G_0$,
shown by dashed green line in the upper subfigure of \fig{GE_W},
indicates the presence of both $0_{+1}$ and $0_{+2}$ incident modes
in the lowest subband $n=0$ caused by the degenerate subband bottom
with finite $q$.  With a window coupling, the transport features of
the $0_{+1}$ and $0_{+2}$ modes are affected.  Increasing the
incident energy, the $0_{+2}$ mode is pushed away the central
barrier by the Lorentz force.  Differently, the $0_{+1}$ mode
demonstrates an electron-like propagation in the low kinetic energy
regime, whereas it exhibits hole-like transport behavior in the high
kinetic energy regime. Both propagation types are steered by the
Lorentz force.  Moreover, conductance oscillations are induced for
the incident electron waves with higher energies just below the
lowest subband top $E=2.95$ meV, as is depicted by solid red curve
in the upper subfigure of \fig{GE_W}.  It has been shown that
conductance oscillation could be useful to obtain information about
the amplitude of the nuclear spin polarization.\cite{Nesteroff04}

\begin{figure}[htbq]
      \includegraphics[width=0.45\textwidth,angle=0]{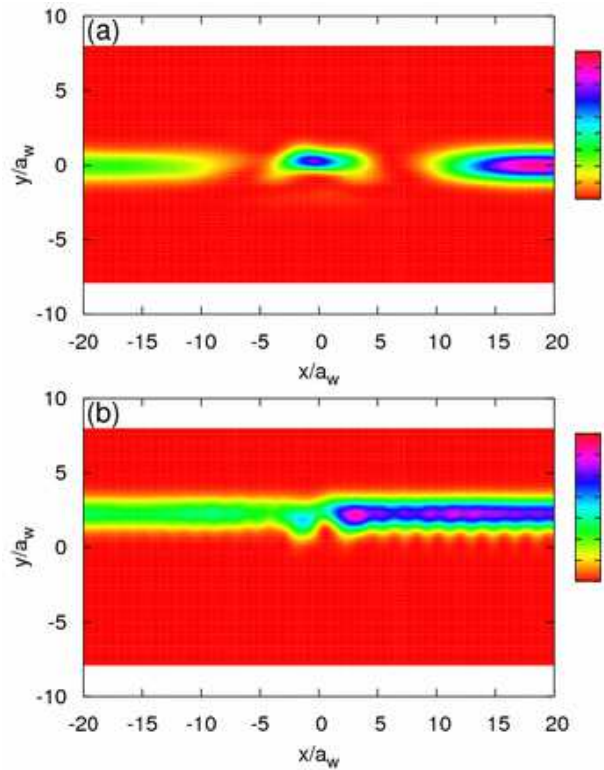}
      \caption{(Color online) The electron probability density at
      $E=2.82$ meV for the $n_i=$ (a) $0_{+1}$ and (b) $0_{+2}$ modes,
      corresponding to the solid red curve in the upper subfigure of \fig{GE_W}.
      $B=0.5$ T and $a_w=29.3$ nm.}
      \label{WfW_E2p8}
\end{figure}

In the low kinetic energy regime, the conductance rises slowly due
to the efficient blocking of the incoming modes.  When the electron
energy is increased, the quantum interference becomes significant,
and hence induces oscillation behavior. The valley structure at
$E=2.64$ meV with $G\approx G_0$ is due to the near total reflection
of the $0_{+1}$ mode around the middle barrier and the resonant
transmission of the $0_{+2}$ outer mode induced by the blocking
potential.  The hump structure in conductance at around the higher
energy $E=2.82$ meV indicates resonant transmission feature of the
two incident modes. Figure \ref{WfW_E2p8}(a) shows that the $0_{+1}$
mode makes only one local resonant state in the window and then
prefers resonant transmission through and above the central region
of the double wire system. Figure \ref{WfW_E2p8}(b) shows that the
outer $0_{+2}$ transport mode prefers transmission slightly
perturbed by the edge blocking potential.  The simple electron
probability feature shown in \fig{WfW_E2p8} implies a negligible
coupling between these two propagating modes.

In the first energy subband gap region $2.95<E<3.11$ meV, the
conductance $G$ is a bit smaller than $G_0$, which implies that only
a single edge transport mode goes through the upper wire, which is
slightly reflected by the edge blocking potential $V_{\rm block}$.
Just above $E=3.11$ meV, the second subband $n=1$ becomes active in
the transport and the total conductance of the system reaches
$1.6G_0$, which is less than $2G_0$ due to the blocking effect of
edge potential.  For electrons with energies $3.24<E<3.42$ meV,
there are four incident modes: $1_{+1}$, $1_{+2}$, $1_{+3}$, and
$0_{+1}$ due to the complicated structure of the second subband
$n=1$. In this energy regime, the highest conductance $G = 3.3G_0$
is at $E=3.32$ meV, which is less than $4G_0$ mainly because of the
strong reflection effect in the coupling window for the $1_{+2}$
mode.

\begin{figure}[htbq]
      \includegraphics[width=0.45\textwidth,angle=0]{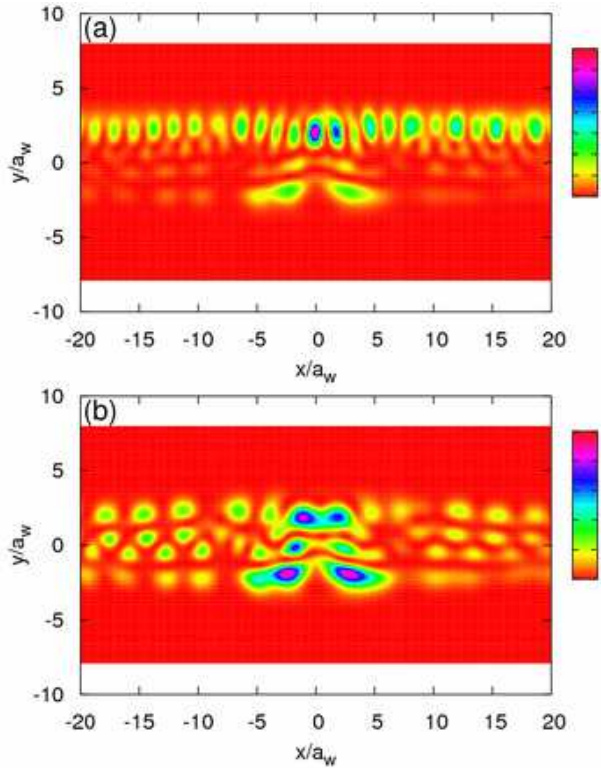}
      \caption{(Color online) The electron probability density at
      $E=3.51$ meV for the $n_i=$ (a) $0_{+1}$ outer and (b) $1_{+1}$ inner modes,
      corresponding to the solid red curve in the upper subfigure of \fig{GE_W}.
      $B=0.5$ T and $a_w=29.3$ nm.}
      \label{WfW_E3p5}
\end{figure}
In the second energy subband gap region $3.42 < E < 3.77$ meV, there
are two incident modes allowed to propagate.  The edge blocking
potential together with the appropriate strength of magnetic field
steers the electron waves to the window region and then enhances
multiple scattering resulting in scarring of wave functions and
interwire transfer effects. The dip structure in conductance at
$E=3.51$ meV demonstrates clearly such interesting transport
dynamics. Figure \ref{WfW_E3p5}(a) shows that the $0_{+1}$ outer
mode propagates along the edge of the upper wire and the coupling to
the lower wire is visible.  Interestingly, \fig{WfW_E3p5}(b) shows
that due to the smaller $q$ of the $1_{+1}$ mode, it is able to
reveal stronger interwire transverse resonant features and then
manifest resonant quasibound state with coupling to the lower wire.
We note in passing that the probability density of the lowest
subband $0_{+1}$ mode usually exhibits simple pattern, the complex
probability pattern shown in \fig{WfW_E3p5}(a) implies a mechanism
of inter-mode transition.

Similar features can be found in another dip structure in
conductance at higher energy $E=3.70$ meV. Localized interwire
resonant states can be found for the case of window coupling. To
demonstrate the possibility of extended interwire resonant transfer,
below we shall discuss the case when the transfer window is embedded
with a transversely coupled resonator.

\subsection{Resonator coupling}

In order to improve the interwire coupling, we investigate the
window resonator coupled transport characteristics of the
edge-blocked double wire. To this end, we select the physical
parameters $V_0 = -2 V_{d_0}$, $\alpha_{\rm R} =0.2\alpha_{\rm W}$,
and $\beta = 0.05\beta_0$ to form a deeper and broader Gaussian
potential and then create a longer window $L_{\rm W}=158$ nm with a
transversely coupled double open-dot resonator, as illustrated in
\fig{System}(b). The longer window can more efficiently interfere
with the wave with the help of the Lorentz force.  The strength of
the edge blocking potential is $V_1 = 8$ meV.  Below we shall
demonstrate coherent interwire magnetotransport features for
different values of magnetic field for comparison.

\subsubsection{$B=0.5$ ${\rm T}$}

\begin{figure}[tbhq]
      \includegraphics[width=0.45\textwidth,angle=0]{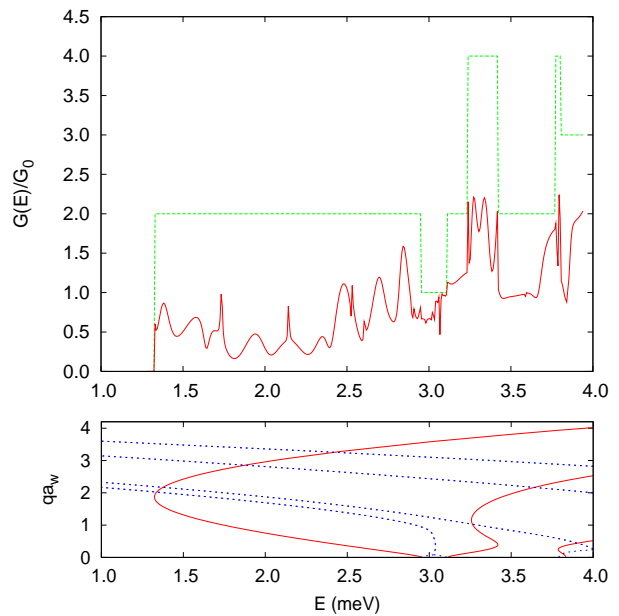}
      \caption{(Color online) Upper subfigure: Conductance of an
      edge-blocked double quantum wire without
      coupling element (dashed green) and with a resonator coupling between
      the wires (solid red).
      Lower subfigure: The energy spectrum of the propagating
      electronic states (solid red) vs.\ $q$, and the energy spectrum of the evanescent
      states (dotted blue) vs.\ $iq$,
      in the asymptotic regions of the system. $B=0.5$ T,
      $V_0 = -2 V_{d_0}$, $\alpha_{\rm R} =0.2\alpha_{\rm W}$,
      $\beta = 0.05\beta_0$, and $V_1=8$ meV.}
      \label{GE_WR}
\end{figure}
In the upper subfigure of \fig{GE_WR}, we show the numerical results
for the case of $B=0.5$ T, the window resonator coupling leads to a
richer conductance spectrum. The lowest subband is conducting up to
$E=2.95$ meV with an inner $0_{+1}$ and an outer $0_{+2}$ incoming
transport modes.  In the low kinetic regime, the conductance $G<G_0$
at $E<2.46$ meV implies an efficient blocking of the $0_{+2}$ mode.
The sharp peaks at energies $1.73$ meV ($G\approx G_0$) and $2.14$
meV ($G\approx 0.8G_0$) imply stronger inter-mode mixing. In
addition, both propagating modes can form well defined localized
states due to the transversely coupled resonator in the window.  In
the high kinetic regime, the electrons propagate demonstrating
different dynamics: The conductance oscillation peaks have values
$G>G_0$ implying that one of the active modes prefers forward
transmission made possible by interwire forward transfer.

In the second subband gap region $3.42<E<3.77$ meV, when the
in-state energy is below $3.67$ meV, the $0_{+1}$ outer  mode is
well resisted by the edge-blocking potential. Hence, the conductance
is decreased to $G \approx G_0$. For in-state energy higher than
$3.67$ meV, the electronic conductance is increased approaching
$2G_0$. This means that the outer mode has sufficient kinetic energy
to pass the edge-blocking potential.

\begin{figure}[htbq]
      \includegraphics[width=0.45\textwidth,angle=0]{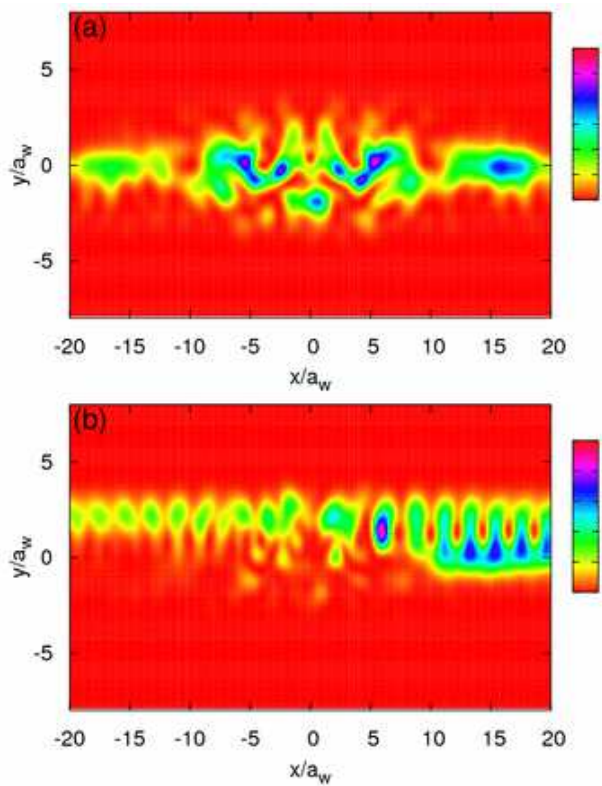}
      \caption{(Color online) The electron probability density at
      $E=2.48$ meV for the $n_i=$ (a) $0_{+1}$ and (b) $1_{+1}$ modes,
      corresponding to the solid red curve in the upper subfigure of \fig{GE_W}.
      $B=0.5$ T and $a_w=29.3$ nm.}
      \label{WfR_E2p4}
\end{figure}
In \fig{WfR_E2p4}(a), we show the electron probability density for
the $0_{+1}$ transport mode, which propagates close to the central
barrier manifesting strong multiple scattering and performing a
symmetric quasibound-state pattern along the transport direction. On
the other hand, the outer $0_{+2}$ transport mode, shown in
\fig{WfR_E2p4}(b), has higher energy to perform resonant
transmission to pass the edge blocking and then make more visible
interwire coupling in the lower right lead.  However, the
entanglement between the propagating modes in the upper and lower
wires dominates the transport feature.

\subsubsection{$B=0.8$ ${\rm T}$}

In order to improve interwire transfer, we select a stronger
magnetic field $B=0.8$ T such that the effective magnetic
confinement length $a_w=25.8$ nm. The accuracy of these high
magnetic field results has been checked by increasing all relevant
grid or basis set sizes used in the numerical calculation. The other
physical parameters remain the same: $V_0 = -2 V_{d_0}$,
$\alpha_{\rm R} =0.2\alpha_{\rm W}$, $\beta = 0.05\beta_0$, and the
strength of the edge blocking potential $V_1 = 8$ meV.

\begin{figure}[tbhq]
      \includegraphics[width=0.45\textwidth,angle=0]{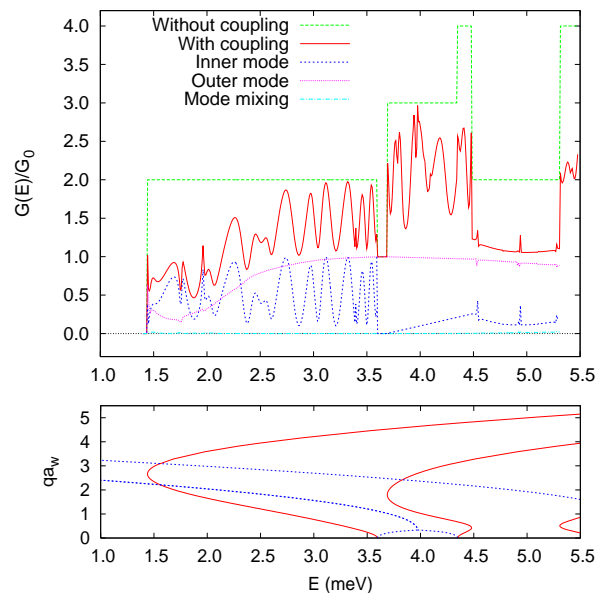}
      \caption{(Color online) Upper subfigure: Conductance of an
      edge-blocked double wire without
      coupling element (dashed green) and with a resonator coupling between
      the wires (solid red); and transmission probabilities of
      inner mode (dashed blue), outer mode (dotted peach),
      and the mode mixing (dash-dotted watchet) of the two modes.
      Lower subfigure: The energy spectrum of the propagating
      electronic states (solid red) vs.\ $q$, and the energy spectrum of the evanescent
      states (dotted blue) vs.\ $iq$,
      in the asymptotic regions of the system. $B=0.8$ T,
      other parameters are in the text.}
      \label{GE_B08_WR}
\end{figure}
The conductance for the case of $B=0.8$ T is shown by solid red
curve in the upper subfigure of \fig{GE_B08_WR}. In comparison with
the lower subfigure of \fig{GE_B08_WR}, we see that the conduction
electrons in the lowest subband at the energy region $1.44<E<3.60$
meV exhibit higher transmission than the case of $B=0.5$ T. In this
energy region, the transmission of the $0_{+2}$ outer mode increases
almost monotonically indicating its increasing ability to pass the
edge-blocking potential, as is illustrated by the dotted peach curve
in the upper subfigure of \fig{GE_B08_WR}. More interestingly, the
strong conductance oscillation of the $0_{+1}$ inner mode (dashed
blue curve) implies rich and energy sensitive resonant transport
features and provides the possibility of efficient interwire
transfer. The mode mixing of the two lowest modes shown by the
dash-dotted watchet curve is generally not active but is slightly
perturbed for the energy $E>4.5$ meV.  For the first subband gap
region at $3.60<E<3.69$ meV, the ideal behavior of conductance
$G=G_0$ indicates the perfect transmission of the $0_{+1}$ outer
mode away from the central barrier.

\begin{figure}[tbhq]
      \includegraphics[width=0.45\textwidth,angle=0]{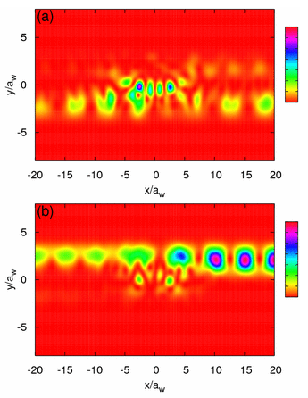}
      \caption{(Color online) The electron probability density at
      $E=1.69$ meV for the $n_i=$ (a) $0_{+1}$ inner and (b) $0_{+2}$ outer modes,
      corresponding to the solid red curve in the upper subfigure of \fig{GE_W}.
      $B=0.8$ T such that $a_w=25.8$ nm.}
      \label{WfR_B08_E1p7}
\end{figure}
The electron probability density shown in \fig{WfR_B08_E1p7}
corresponds to the electron modes at incident energy $E=1.69$ meV
with conductance maximum $G\approx 0.9 G_0$. Figure
\ref{WfR_B08_E1p7}(a) demonstrates the transport properties of the
$0_{+1}$ low $q$ mode incident from the left lower channel.  The
electrons are steered by the Lorentz force into the resonator
coupling region and exhibit resonant features in the window.  It
turns out that the Lorentz force fits the window size and provides
efficient electron {\it interwire forward transfer} to the right
upper channel.  In \fig{WfR_B08_E1p7}(b), we see that the $0_{+2}$
high $q$ incident mode is strongly affected by the Lorentz force,
and thus incident from the left upper channel.  The electrons
entering the resonator coupling element region exhibit weak coupling
to the lower wire.  On the other hand, the electrons also perform
resonant transmission through the upper blocking potential with more
completed cyclotron motion. The periodic peak features of the
probability density in the right upper channel indicates a signature
of electron propagation in the sufficient low kinetic energy regime
with strong multiple scattering in the transverse direction.

\begin{figure}[tbhq]
      \includegraphics[width=0.45\textwidth,angle=0]{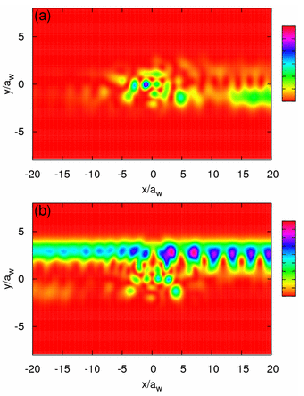}
      \caption{(Color online) The electron probability density at
      $E=2.07$ meV for the $n_i=$ (a) $0_{+1}$ inner and (b) $0_{+2}$ outer modes,
      corresponding to the solid red curve in the upper subfigure of \fig{GE_W}.
      $B=0.8$ T such that $a_w=25.8$ nm.}
      \label{WfR_B08_E2p0}
\end{figure}
In comparison with the conductance maximum at incident energy
$E=1.69$ meV, we  see the case of a little higher incident energy
$E=2.07$ meV with conductance minimum $G\approx 0.6 G_0$, the
corresponding probability density is shown in \fig{WfR_B08_E2p0}.
Figure \ref{WfR_B08_E2p0}(a) demonstrates the transport property of
the $0_{+1}$ mode.  We see that the increasing of the kinetic energy
suppresses a little the interwire forward transfer with negligible
mode mixing.  In \fig{WfR_B08_E2p0}(b), we show the probability
density of the $0_{+2}$ mode incident from the left upper channel.
This outer mode makes stronger resonant transmission in the upper
wire, and the interwire coupling is enhanced to exhibit interwire
backward transfer to the left lower channel.  We note in passing
that even for such a simple transfer mechanism, all possible
intermediate states have to be taken into account in the
calculation, which can not be obtained using a low order
perturbation theory.\cite{Yang91}

It should be noticed that when the electron Fermi energy is above
the lowest subband local top $3.60$ meV, the lowest active inner
mode is changed to be $1_{+1}$ mode, and the lowest outer mode would
be $0_{+1}$ mode. For the second subband gap region $4.48<E<5.31$
meV, the conductance plateau $G \leq G_0$ indicates that the
$1_{+1}$ inner mode is efficiently backscattered by the window
resonator. In addition, the $0_{+1}$ outer mode manifests near ideal
transmission due to its high kinetic energy.

We would like to mention that only the larger $q$ evanescent modes
exist in the second gap region.  This fact leads to stronger
localized bound states in the window resonator with longer dwell
time. Especially, there are three small conductance peaks at
energies $E=4.54$, $4.94$ and $5.28$ meV in the second subband gap
region. By checking the inner mode and the outer mode contributions
to the conductance shown in the upper subfigure of \fig{GE_B08_WR},
we see that both of them manifest small sharp change in conductance
at these energies. We have tested the case of $E=4.94$ meV to see
that both modes exhibit amazingly similar scarring resonant state
patterns in the window covering the upper and the lower wire, which
is an example of persistence of a scarring wave function in an open
system.\cite{Akis02}

%
\section{concluding remarks}

In this report we have investigated theoretically to what extend the
coherent magnetotransport properties in an edge-blocked lateral
double wire system with possible resonant interwire coupling in the
presence of a uniform perpendicular magnetic field.  The complex
subband structure, due to the non-parabolic double wire confinement
in the magnetic field, causes the appearance of irregular steps in
the conductance as a function of the energy of the incoming electron
wave.  Even with the coupling element in the tunneling regime, a
usual perturbation theory is not valid once the length of the
coupling region exceeds a characteristic length scale.\cite{Boese01}
Our numerical method employed allows for a wide variety of wire
shape and scattering potentials.

To enhance the interwire forward or backward transport, we have
shown that not only appropriate window size but transversely coupled
resonator and the proper magnetic field strength are required. The
electron probability density shown in this report should be
detectable by using high-resolution scanning-probe
images.\cite{Crook_N03,Crook_L03,Mendoza06}  We have also
demonstrated that the mode mixing in energy regions containing two
active modes is relatively weak, we expect that the conductance
oscillations could be clearly observable.  An efficient interwire
forward transfer can be achieved by the inner mode with negligible
mode mixing in a properly applied magnetic field, whereas the outer
mode exhibits interwire backward transfer.  The magnetic field
manipulated window and resonator coupling features in a double wire
system are important to understanding magnetotransport properties of
other open quantum structures.


%
\begin{acknowledgments}
      The authors acknowledge the financial support by the Research
      and Instruments Funds of the Icelandic State,
      the Research Fund of the University of Iceland, and the
      National Science Council of Taiwan.
      C.S.T. is grateful to inspiring discussions with P.G. Luan and S.A. Gurvitz, and
      the computational facility supported by the
      National Center for High-Performance Computing of Taiwan.
\end{acknowledgments}

%
%

%
%

\begin{thebibliography}{99}


\bibitem{Eugster91} C.~C. Eugster and J.~A. del Alamo, Phys. Rev. Lett.
{\bf 67}, 3586 (1991).

\bibitem{Hirayama92} Y. Hirayama, A.~D. Wieck, T. Bever, K. von Klitzing, and K.
Ploog, Phys. Rev. B {\bf 46}, 4035 (1992).

\bibitem{Xu95} G. Xu, L. Jiang, P. Jiang, D. Lu, and X. Xie, Phys.
Rev. B {\bf 51} 2287 (1995).

\bibitem{Governale00} M. Governale, M. Macucci, and B. Pellegrini,
Phys. Rev. B {\bf 62}, 4557 (2000).

\bibitem{Guzenko06} V.~A. Guzenko, J. Knobbe, H. Hardtdegen, and Th.
Sch{\"a}pers, Appl. Phys. Lett. {\bf 88}, 032102 (2006).

\bibitem{Tang05} C.-S. Tang, W.~W. Yu, and V. Gudmundsson, Phys.
Rev. B {\bf 72}, 195331 (2005).

\bibitem{Gudmundsson06} V. Gudmundsson and C.-S. Tang,
cond-mat/0606480 (unpublished).


\bibitem{Barbosa97} J.~C. Barbosa and P.~N. Butcher, Superlatt.
Microstruct. {\bf 22}, 325 (1997).


\bibitem{Korepov02} S.~V. Korepov and M.~A. Liberman, Physica B {\bf 109},
92 (2002).

\bibitem{Arapan03} S.~C. Arapan, S.~V. Korepov, M.~A. Liberman, and
B. Johansson, Phys. Rev. B {\bf 67}, 115328 (2003).

\bibitem{Fischer05} S.~F. Fischer, G. Apetrii, U. Kunze, D. Schuh,
and G. Abstreiter, Phys. Rev. B {\bf 71}, 195330 (2005); {\it
ibid}., Nature Phys. {\bf 2}, 91 (2006).

\bibitem{Shi97} J.-R. Shi and B.-Y. Gu, Phys. Rev. B {\bf 55}, 9941
(1997).

\bibitem{Gurvitz95} S.~A. Gurvitz, Phys. Rev. B {\bf 51}, 7123 (1995).

\bibitem{Gudmundsson05} V. Gudmundsson, Y.-Y. Lin, C.-S. Tang, V.
Moldoveanu, J.~H. Bardarson, and A. Manolescu, Phys. Rev. B {\bf
71}, 235302 (2005).

\bibitem{Sasaki06} S. Sasaki, S. Kang, K. Kitagawa,
M. Yamaguchi, S. Miyashita, T. Maruyama, H. Tamura, T. Akazaki, Y.
Hirayama, and H. Takayanagi, Phys. Rev. B {\bf 73}, 161303(R)
(2006).


\bibitem{Landauer57} R. Landauer, IBM J. Res. Dev. {\bf 1}, 223
(1957).

\bibitem{Buttiker86} M. B\"{u}ttiker, Phys. Rev. Lett. {\bf 57}, 1761
(1986); {\it ibid}., IBM J. Res. Dev. {\bf 32}, 306 (1988).

\bibitem{Nesteroff04} J.~A. Nesteroff, Y.~V. Pershin, and V.
Privman, Phys. Rev. Lett. {\bf 93}, 126601 (2004).

\bibitem{Yang91} R.~Q. Yang and J.~M. Xu, Phys. Rev. B {\bf 43},
1699 (1991).

 \bibitem{Akis02} R. Akis, J.~P. Bird, and D.~K. Ferry, Appl. Phys. Lett. {\bf
 81}, 129 (2002); Y.-H. Kim, M. Barth, U. Kuhl, H.-J. St{\"o}ckmann, and J.~P.
 Bird, Phys. Rev. B {\bf 68}, 045315 (2003).

\bibitem{Boese01} D. Boese, M. Governale, A. Rosch, and
U. Z{\"u}licke, Phys. Rev. B {\bf 64}, 085315 (2001).

\bibitem{Crook_N03} R. Crook, A.~C. Graham, C.~G. Smith, I. Farrer, H.~E.
Beere, and D.~A. Ritchie, Nature {\bf 424}, 751 (2003).

\bibitem{Crook_L03} R. Crook, C.~G. Smith, A.~C. Graham, I. Farrer, H.~E.
Beere, and D.~A. Ritchie, Phys. Rev. Lett. {\bf 91}, 246803 (2003).

 \bibitem{Mendoza06} M. Mendoza and P.~A. Schulz, Phys. Rev. B {\bf
 74}, 035304 (2006).

\end{thebibliography}
\end{document}